\documentstyle[psfig]{lamuphys}

\def\amin{\ifmmode ^{\prime}\else$^{\prime}$\fi}
\def\asec{\ifmmode ^{\prime\prime}\else$^{\prime\prime}$\fi}
\def\x{\mbox{\boldmath $\xi$}}
\def\vi{\hat{\mbox{\boldmath $\rm i$}}}
\def\vj{\hat{\mbox{\boldmath $\rm j$}}}
\def\vk{\hat{\mbox{\boldmath $\rm k$}}}
\def\vphi{\hat{\mbox{\boldmath $\varphi$}}}
\def\nab{\mbox{\boldmath $\nabla$}}

\begin{document}        % ignore

\title{Precessing warped discs in close binary systems}

% If your title plus authors is too long for one line in the automatic
% page heading, you have to supply a short version of the title in the
% following command.
\titlerunning{Precessing discs}

% If there are no co-authors, you can delete the \inst{} commands.
\author{J.C.B. Papaloizou \inst{1}, J.D. Larwood \inst{1}, R.P. Nelson\inst{1}, C. Terquem \inst{1,2}}

\institute{ Astronomy Unit, School of Mathematical Sciences, Queen Mary and
Westfield College, Mile End Road, London E1 4NS
\and
Laboratoire d'Astrophysique, Observatoire de Grenoble,
Universit\'e Joseph Fourier/CNRS, BP 53X, 38041 Grenoble Cedex,
France}

\maketitle

% The following \index commands are for the author index. The first author
% gets a "|uu" appended, all co-authors the simple format. Please, substitute
% your name here:
\index{Author, J.B.C. Papaloizou |uu}
\index{Buthor, J.D. Larwood}
\index{Cuthor, R.P. Nelson}
\index{Duthor, C. Terquem}

% For the subject and object index you can distribute \glossary{} commands
% throughout your contribution. General keywords for your article which
% should appear only once (with the first page of your article) in the index,
% can be given here. The remaining should come in the text after the appearance
% of the appropriate words/phrases.
\glossary{accretion discs}
\glossary{Close binary systems}
\glossary{Tidal effects}
\glossary{Precessing discs}
\glossary{Warped discs}

% In general, short papers (up to 4 pages) should not have an abstract. Please,
% delete the following lines in this case.
\noindent{\bf Abstract:}
We describe some recent nonlinear three dimensional hydrodynamic simulations of
accretion discs in binary systems where the orbit is circular and not necessarily
coplanar with the disc midplane.
 
\noindent The calculations are relevant to a number of observed astrophysical
phenomena,
including the precession of jets associated with young stars, the high
spectral index of some T~Tauri stars, and the light curves of X--ray
binaries such as Hercules X-1 which suggest the presence of precessing
accretion discs.
 
\section{Introduction}
 
Protostellar discs appear to be common around young stars. Furthermore
recent studies show that almost all young stars associated with low
mass star forming regions are in multiple systems (Mathieu, 1994 and
references therein). Typical orbital separations are around 30
astronomical units (Leinert et al. 1993) which is smaller than the
characteristic disc size observed in these systems (Edwards et
al. 1987). It is therefore expected that circumstellar discs will be
subject to strong tidal effects due to the influence of binary
companions.
 
\noindent The tidal effect of an orbiting body on a differentially rotating disc
has been well studied in the context of planetary rings (Goldreich and
Tremaine, 1981), planetary formation, and generally interacting binary
stars (see  Lin and  Papaloizou, 1993 and references therein). In these
studies, the disc and orbit are usually taken to be coplanar (see
Artymowicz and  Lubow, 1994). However, there are observational indications
 that  discs and stellar orbits may not always be coplanar ( see for example
 Corporon, Lagrange and  Beust, 1996 and  Bibo, The and  Dawanas, 1992.)
 
\noindent  In addition, reprocessing of radiation from the central star by a
warped non coplanar disc has been suggested in order to account for the high
spectral index of some T~Tauri stars (Terquem and Bertout 1993, 1996).
 
\noindent A dynamical study of the tidal interactions of a non coplanar disc is
of interest not only in the above contexts, but also in
relation to the possible existence of precessing discs which may define
the axes for  observed jets which apparently precess (Bally and Devine, 1994).
 
\noindent Various studies  of the evolution of warped  discs have been undertaken
assuming that the forces producing the warping were small so that
linear perturbation theory could be used ( Papaloizou and Pringle, 1983,
Papaloizou and  Lin, 1995 and  Papaloizou and  Terquem, 1995). The
results suggested that the disc would precess approximately as a rigid body if
the sound crossing time was smaller than the differential precession frequency.
 
We describe here some recent non linear simulations of
 discs which are not coplanar
with the binary orbits using a Smoothed Particle
Hydrodynamics (SPH) code originally developed by Nelson and
Papaloizou (1993, 1994).  We  study the conditions under which
  warped precesing discs may survive in close binary
systems and  the truncation of the disc size through
tidal effects when the disc and binary orbit are not
coplanar. The simulations indicate  that the phenomenon of tidal truncation
is only marginally affected by lack of coplanarity. Also our model discs were
able to survive in a tidally truncated condition while warped and undergoing
rigid body precession provided that the
 Mach number in the disc was not too large. The
inclination of the disc was found to evolve  on a long timescale likely to
be the viscous timescale, as was indicated by the linear calculations of
Papaloizou and  Terquem (1995).

\section{Basic equations}
 
The equations of continuity and of motion applicable to a gaseous
viscous disc may be written

\begin{equation}  
\frac{d\rho}{dt} + \rho \nab \cdot {\bf v} = 0 ,
\label{cont} 
\end{equation}

\begin{equation}
\frac{d{\bf v}}{dt} = - \frac{1}{\rho} \nab P - \nab \Psi +
{\bf{S}}_{visc} 
\label{dvdt} 
\end{equation}

\noindent where 

\begin{displaymath}
{d \over dt} \equiv {\partial \over \partial t} +
{\bf v} \cdot \nab
\end{displaymath}

\noindent denotes the convective derivative, $\rho$ is the density, 
${\bf v}$ the velocity and $P$ the pressure. The gravitational
potential is $\Psi,$ and ${\bf{S}}_{visc}$ is the viscous force per
unit mass.

\noindent For the work described here, we adopt the polytropic 
equation of state

\begin{displaymath}
P= K \rho^{\gamma}
\end{displaymath}

\noindent where

\begin{displaymath}
c_s^2 = K \gamma \rho^{\gamma - 1}
\end{displaymath}

\noindent gives the usual associated sound speed, $c_s$. Here we take 
$\gamma=5/3,$ and $K$ is the polytropic constant. This corresponds to
adopting a fluid that remains isentropic throughout even though
viscous dissipation may occur. This means that an efficient cooling
mechanism is assumed.

\section{Orbital Configuration}

We consider a binary system in which the primary has a mass M$_p$ and
the secondary has a mass M$_s.$ The binary orbit is circular with
separation $D.$ The orbital angular velocity is $\omega.$ We suppose
that a disc orbits about the primary such that at time $t=0$ it has a
well defined mid-plane. We adopt a non rotating Cartesian coordinate
system $(x,y,z)$ centred on the primary star and we denote the unit
vectors in each of the coordinate directions by $\vi$, $\vj$ and $\vk$
respectively. The $z$ axis is chosen to be normal to the initial disc
mid-plane. We shall also use the associated cylindrical polar
coordinates $(r,\varphi,z).$

\noindent We take the orbit of the secondary to be in a plane which has an initial
inclination angle $\delta$ with respect to the $(x,y)$ plane. For a
disc of negligible mass, the plane of the orbit is invariable and does
not precess. We denote the position vector of the secondary star by
$\bf{D}$ with $D=|{\bf D}|.$ Adopting an orientation of coordinates
and an origin of time such that the line of nodes coincides with, and
the secondary is on, the $x$ axis at $t=0,$ the vector ${\bf D}$ is
given as a function of time by

\begin{equation}
{\bf D} = D \cos \omega t \;  \vi +
D \sin \omega t \cos \delta  \; \vj +
D \sin \omega t \sin \delta \; \vk .
\label{x}
\end{equation}

\noindent The total gravitational potential $\Psi_{ext}$ due to the binary pair at a point
with position vector ${\bf r}$ is given by

\begin{displaymath}
\Psi_{ext} = - \frac{GM_p}{\mid \bf{r} \mid} -
\frac{GM_s}{\mid \bf{r} - \bf{D} \mid} +
{GM_s{\bf r} \cdot {\bf D}\over D ^3}
\end{displaymath}

\noindent where $G$ is the gravitational constant. The first dominant 
term is due to the primary, while the remainder, $\equiv \Psi_{ext}',$
 gives perturbing terms due to the secondary. Of these, the last indirect term
accounts for the acceleration of the origin of the coordinate system. We note
that a disc perturbed by a secondary on an inclined orbit becomes tilted,
precesses and so does not maintain a fixed plane. Our calculations
presented below are referred to the Cartesian system defined above
through the initial disc mid-plane. However, we shall also use a
system defined relative to the fixed orbital plane for which the `$x$
axis' is as in the previous system and the `$z$ axis' is normal to
the orbital plane. If the disc were a rigid body its angular momentum
vector would precess uniformly about this normal, as indicated below.
\subsection{Disc response}
The form of the disc response to the perturbing gravitational potential due to
the secondary is determined by the properties of the free modes of oscillation.
These are divided into two classes according to whether the associated density
perturbation has even or odd symmetry with respect to reflection in the
unperturbed disc midplane. The modes with even symmetry are excited when
the binary orbit and disc are coplanar and have been well studied in the
context of angular momentum exchange between disc and binary leading
to tidal truncation of and wave excitation in the disc ( see Lin and Papaloizou,
1993 and references therein). They are also excited at a somewhat reduced
level in the non coplanar case where they produce similar effects.

\noindent Here we shall focus attention on the modes with odd symmetry
and with azimuthal mode number $m=1.$
These are only excited in the non coplanar case. They are of interest
because they are responsible for disc warping, twisting and precession.

\subsection{Potential expansion}
When the orbital separation is much larger then the outer disc radius such that
for any disc particle, $r \ll D,$ and  $z \ll D,$ 
we can expand $\Psi_{ext}'$ in powers of $r/D$ and $z/D$.   

\noindent We are interested in bending modes which are excited by terms in the potential
which are odd in $z$ and which have azimuthal mode number $m=1$ when a Fourier
analysis in $\varphi$ is carried out. The lowest order terms in the expansion
of the  potential which are of the required form are given by

\begin{eqnarray}
\Psi'_{ext} = & - \frac{3}{4} \frac{GM_s}{D^3} r z & \left[
\left( 1 - \cos \delta \right)
\sin \delta \sin \left( \varphi + 2 \omega t \right) \right.
\nonumber \\
 & & \left. - \left( 1 + \cos \delta \right) \sin \delta
\sin \left( \varphi - 2 \omega t \right) \right.
\nonumber \\
 & & \left. + \sin 2 \delta
\sin \left( \varphi  \right) \right]
\label{psi'}
\end{eqnarray}

\noindent In  linear perturbation theory, we can calculate  the
response of the disc to each of the three terms in $\Psi'_{ext}$ separately and
superpose the  results. The general problem is then to calculate the response
due to a potential of the  form

\begin{displaymath}
\Psi'_{ext} = {\rm const}\times r z \sin \left( \varphi - \Omega_P t \right)
\end{displaymath}

\noindent or in complex notation

\begin{equation}
 {\Psi'_{ext}} = {\cal{R}}\left(f r z e^{i m\left( \varphi - \Omega_P t
\right)}\right)\label{lpot} \end{equation}

\noindent  Here, the azimuthal mode number, $m=1$,
 the pattern frequency $\Omega_P$
of the perturber is one of $0,$  $2 \omega $ or $- 2 \omega $ and
$f$ is the appropriate complex amplitude. We remark that the magnitude of the
tidal perturbation acting on the disc is measured by the dimensionless
quantity $GM_s/(\Omega^2 D^3),\ \Omega$ being the angular velocity in the disc
and for comparable primary and secondary masses this is of order
$\omega^2/\Omega^2.$ 

\section{Free bending modes}
Bending waves with $m=1$ are naturally excited by a perturbing potential
of the form given by $(\ref{lpot}).$ For a binary with large
orbital radius, we may consider the pattern frequency
$\Omega_P$ to be small compared to the rotation frequency in the disc, $\Omega.$
Thus the forcing is at low frequency.

\noindent Bending waves in thin discs have been studied in the context of disc
galaxies (Hunter and Toomre, 1969), planetary rings (Shu,1984) and gaseous
accretion discs ( Papaloizou and Lin, 1995, Papaloizou and Terquem, 1995). 

\noindent In a self-gravitating disc with no pressure and of small enough mass
that the unperturbed disc is in a state of near Keplerian rotation,
the local dispersion relation for bending waves with $m=1$ is given by (Hunter
and Toomre, 1969)
 \begin{equation} (\Omega_P-\Omega)^2 = \Omega^2 +2\pi G\Sigma |k|,
\end{equation}
where $\Sigma$ is the disc surface density. In the limit of small pattern
speeds this takes the form
  \begin{equation} \Omega_P\Omega = -\pi G\Sigma |k|,
\end{equation} from which it follows that
the waves propagate without dispersion with speed
 \begin{equation}c_g ={\pi G\Sigma \over \Omega}= {\langle c_s\rangle
\over Q},\end{equation}
where the Toomre $Q=\Omega{\langle c_s\rangle }/(\pi G\Sigma),$ 
where the angled brackets denote an apropriate vertical mean of the sound speed.
For stability to axisymmetric modes, we require $Q \ge 1,$ which implies that
bending waves in a stable disc propagate with a speed not exceeding the maximum  sound
speed at a particular radial location.

\noindent  Papaloizou and Lin (1995)
considered the case when pressure is included, giving the corresponding  local
dispersion relation for bending waves with $m=1$ in the low frequency limit 
   \begin{equation} \Omega_P\Omega =-\pi G\Sigma |k|-{(1-\Theta)\Omega \langle c_s\rangle ^2
|k|^2\over 4\Omega_P}, \end{equation}
where $\Theta < 1$ is a dimensionless parameter which vanishes when self-gravity
is unimportant.
 This gives a quadratic equation for $\Omega_P$ with the two roots
 \begin{equation}\Omega_P = - {\langle c_s\rangle |k|\over 2}\left({1\over Q}{\pm}\sqrt{{1\over
Q^2}+1-\Theta}\right). \end{equation}
In this case there are fast and slow  waves which in the case of a stable disc
propagate with speeds comparable to the sound speed. When self-gravity is
unimportant there is a single sonic like wave. The excitation by the
forcing potential (\ref{lpot}) , and
angular momentum transport associated with these waves with non zero
$\Omega_P$ has been
considered by Papaloizou and Terquem (1995).

\noindent The secular term in the forcing potential (\ref {psi'}) with zero
pattern speed  causes the disc to be  subject to a precessional torque. The
properties of the bending waves determine the form of the disc response. For the
disc to precess approximately like a rigid body, the effects of the precessional
torque which acts largely in the outer parts of the disc,  must be communicated
to the inner regions where it is weakest, within a precession period. This
roughly corresponds to the condition that the disc sound crossing time be less
than the precession period.

\section{Precession Frequency}

\noindent In order to calculate the precession frequency, 
we  consider the time independent, or
 secular term in the perturbing potential (\ref {psi'}) as it is only
this term which produces a non zero net torque after performing a 
time average.  This is given by  
\begin{equation}
{\Psi_{ext0}'} = -\frac{3}{4} \frac{GM_s}{D^3} r z 
\sin 2 \delta \sin(\varphi).
\label{psib}
\end{equation}

\noindent  For a conservative system, the Lagrangian displacement vector $\x$
will satisfy an equation of the form (see Lynden-Bell and Ostriker~1967)

\begin{equation}
{\bf C}(\x) = -\nab {\Psi_{ext0}'}. 
\label{RESP} 
\end{equation} 

\noindent Here ${\bf C}$ is a linear operator, which needs to be 
inverted to give the response. When a barotropic equation of state
applies, and the boundaries are free, ${\bf C}$ is self-adjoint with
weight $\rho.$ This means that for two general displacement vectors
$\x({\bf r})$ and $\mbox{\boldmath $\eta$}({\bf r})$ we have
 
\begin{equation}
\int_V \rho \mbox{\boldmath $\eta$}^* \cdot
{\bf {C}} \left( \x \right) d\tau =
\left[ \int_V \rho \x^* \cdot {\bf {C}} \left(
\mbox{\boldmath $\eta$} \right) d\tau \right]^*
\end{equation}
 
\noindent where $^*$ denotes complex conjugate and the integral is 
taken over the disc volume V.
 
\noindent Because of the spherical symmetry of the
unperturbed primary potential, the unperturbed system is invariant under applying
a rigid tilt to the disc. This corresponds to the existence of rigid tilt mode
solutions to (\ref{RESP}) when there is no forcing ($\Psi'_{ext0}=0)$. For a
rotation about the $x$ axis, the rigid tilt mode is of the form

\begin{equation} 
\x = \x_{T} =
\left( {\bf r} \mbox{\boldmath $\times$} 
\vphi \right) \sin \varphi - \vphi z \cos \varphi
\label{tilt}
\end{equation} 

\noindent where $\vphi$ is the unit vector along the $\varphi$ 
direction. For time averaged forcing potentials $\propto \sin\varphi,$
the existence of the solution~(\ref{tilt}) results in an integrability
condition for (\ref{RESP}). When ${\bf C}$ is self-adjoint, this is

\begin{equation} 
\int_V \x_T \cdot \nab {\overline \Psi_s} 
\rho d\tau \equiv \int_V \vi \cdot
\left ({\bf r} \mbox{\boldmath $\times$} \nab 
{\overline \Psi_s}\right) \rho d\tau = 0. 
\label{INTCON} 
\end {equation}

\noindent The above condition is equivalent to the requirement that the $x$
component of the external torque vanishes. This will clearly not be
satisfied in the problem we consider.

\noindent To deal with this one may suppose that the disc angular momentum
vector precesses about the orbital angular momentum vector with
angular velocity $\mbox{\boldmath $\omega_p$}$ (Papaloizou and
Terquem~1995). This in turn is equivalent to supposing our coordinate
system rotates with angular velocity $\mbox{\boldmath $\omega_p$}$
about the orbital rotation axis. Treating the  Coriolis
force by perturbation theory produces an additional term on the right hand side
of~(\ref{RESP}) equal to 
$-2 r\Omega \mbox{\boldmath $\omega_p$} \mbox{\boldmath
$\times$} {\hat{\mbox{\boldmath $\varphi$}}}$. Using the modified force in
formulating~(\ref{INTCON}) gives the integrability condition as

\begin{equation} 
\omega_p \sin \delta \int_V r^2 \Omega \rho d\tau =
\int_V \vi \cdot \left ({\bf r} \mbox{\boldmath $\times$} 
\nab {\overline \Psi_s} \right) \rho d\tau , 
\label{PRECESS} 
\end{equation}

\noindent with $\omega_p = |\mbox{\boldmath $\omega_p$}|.$ 
Equation~(\ref{PRECESS}) gives a precession frequency for the disc
that would apply if it were a rigid body. However, approximate rigid
body precession is only expected to occur if the disc is able to
communicate with itself, either through wave propagation or viscous
diffusion, on a timescale less than the inverse precession frequency
(see for example Papaloizou  and Terquem~1995 and below). Otherwise, a
thin disc configuration may be destroyed by strong warping and
differential precession.

\noindent We comment that the situation described above in which the
external perturbation produces a precessional torque in the $x$
direction only is a consequence of the assumption of a conservative
response for which the density and potential perturbations are
in phase. However, if dissipative processes are included, there will
be a phase shift between the perturbing potential and density response.
This will result in a net torque in the $y$ direction which can change
the angle between the disc and orbital angular momentum vectors
(see Papaloizou and Terquem, 1995) possibly leading to their alignment. 
Such a process is likely to occur on the long dissipative timescale.
A torque in the $y$ direction originating from a disc wind has been
proposed by  Schandl and Meyer (1994) in order to produce misalignment
between the disc and binary orbit angular momentum vectors in HZ-Hercules.

\section{Numerical Simulations}
Three dimensional simulations of warped precessing discs in close binary systems
have been carried out by solving
the  set of basic equations~(\ref{cont}) and~(\ref{dvdt})
numerically using an SPH code
(Lucy~1977, Gingold  and Monaghan~1977), developed by Nelson and 
Papaloizou~(1994), which uses a conservative formulation of the method
that employs variable smoothing lengths. 
A suite of test calculations illustrating the
accurate energy conservation obtained with this method is described by
Nelson  and Papaloizou~(1994), and additional tests and calculations are 
presented in Nelson~(1994).

\noindent Larwood et al (1996) considered circumprimary discs in close
binary systems with mass ratio of order unity and Larwood and Papaloizou (1997)
have considered circumbinary discs in systems with a variety of mass ratios.

\noindent In order to stabilize the calculations in the presence of shocks, 
the artificial viscous pressure prescription of
Monaghan~and Gingold~(1983) has been used in the simulations.
This induces a shear viscosity which 
leads to angular momentum transport and the standard viscous evolution
of an accretion disc (Lynden-Bell  and Pringle~1974) in which disc expansion is
produced by outward transport of angular momentum as mass flows inwards.  In the
studies  presented here, the discs undergo angular momentum loss through
tidal interaction with orbiting secondaries. Then
disc expansion arising from outward transport of angular momentum is halted by
tidal truncation. This effect, well known in the coplanar case ( see for
example Lin~and Papaloizou~1993), also occurs here when the disc and binary
angular momenta are not aligned.

\noindent In the simulations reported here the shear viscosity, $\nu$
operating in our disc models  was  well fitted by a
constant value. To specify the magnitude of $\nu$ we
write $\nu =\alpha c_s^2(R)/\Omega (R),$ where $\alpha$
corresponds to the well known Shakura~and Sunyaev~(1973) $\alpha$
parameterization and $R$ denotes the outer radius of the disc. However, it is
applicable only at the outside edge of the disc. The discs were
here set up with a distribution of 17500 particles such that the surface density
was independent of radius. Then the aspect ratio $H/r,$ with $H$ being the
semi-thickness was found to be approximately independent of radius.
As a consequence of this the Mach number ${\cal M}=(H/R)^{-1}$ is also
approximately constant for a particular simulation and can be used to
parameterize it. Also the radial dependence of the viscosity
parameter is $\alpha \propto r^{-1/2}.$

\noindent We  comment that a characteristic value of $\alpha =0.03$ that we have in the
simulations  is about two
orders of magnitude larger than that expected to be associated with tidally
induced inwardly propagating waves (Spruit~1987, Papaloizou and 
Terquem~1995). Accordingly, it is expected that tidal truncation will
instead occur through strong nonlinear dissipation near the disc's
outer edge (Savonije, Papaloizou  and Lin~1994). The large viscosity
of the disc models considered here will damp inwardly propagating
waves before they can propagate very far.

\noindent In order to deal with the central regions of the disc, the primary's
gravitational potential softened  such that

\begin{displaymath}
\Psi_{p} = - {GM_p \over \sqrt{r^2 +b^2}} ,
\end{displaymath}

\noindent where $b$ is the softening length. 

\noindent  We adopt units such that the
primary mass $M_p=1$, the gravitational constant $G=1$, and the outer disc radius
$R=1$. In these units the adopted softening parameter $b=0.2$ and the time
unit is $\Omega(R)^{-1}.$ The self-gravity of the disc material has been
neglected in the simulations presented here. 

\section{Numerical Results}

We have considered the evolution of  disc models set up according
to the procedure outlined above. The models were
characterized only by the mean Mach number, ${\cal M}$. After a relaxation
period of about two rotation periods at the outer edge of the disc, the time was
reset to zero and the secondary was introduced in an inclined circular orbit,
moving in a direct sense, crossing the $x$ axis at $t=0.$  In some cases the full
secondary mass was included immediately corresponding to a sudden
start. However, for strong initial tidal interactions such as those
that occur when $D/R=3, M_s/M_p=1$, this can result in
disruption of the outer edge of the disc with a small number of particles
being ejected from the disc. It was found that this
could be avoided by using a
`slow start' in which $M_s$ was built up gradually (see Larwood et al,1996).
However, subsequent results were found to be indepent of the initiation
procedure.

\noindent In the above discussion of bending modes we indicated that
the disc should be able to approximately precess as a rigid body if
the sound crossing time was short compared to the characteristic precesion
period. 

\noindent   The general finding from the simulations was that a
disc with an initial angular momentum vector inclined to that of the
binary system tended to precess approximately as a rigid body, with a
noticeable but small warp if ${\cal M}$ was not too large. In such
cases only small changes in the inclination angle between the angular
momentum vectors were found over the run time. This is consistent with
the expectation from Papaloizou  and Terquem~(1995) that the timescale
for evolution of the inclination in such cases should be comparable to
the viscous evolution timescale of the disc, assuming outward disc
expansion is prevented by tidal interaction.

\noindent We here describe simulations of three circumprimary disc models  in
a close binary system with $D/R=3$ and $\delta =\pi /4$ initially.
Models 1 ,2 and 3 had ${\cal M}$ equal to 20, 25 and 30 respectively
and the total run times for these models initiated with a slow start
was 310, 217.7, and 397.9 units respectively. Note that the viscosity
is larger in the models with smaller Mach number so that this aids
disc communication in these cases also.

\noindent The calculations presented here use a coordinate system which is based
on the initial disc mid-plane. However, as the disc precesses, the
mid-plane changes location with time. It is then more convenient to
use a coordinate system $(x,y_o,z_o)$ based on the fixed orbital
plane, the $z_o$ axis coinciding with the orbital rotation axis.
We shall refer to these
as `orbital plane coordinates'. We locate the inclination angle
$\iota$ (equal to $\delta$ at $t=0$) between the disc and binary orbit
angular momentum vectors through

\begin{displaymath}
\cos \iota = { {\bf J}_D\cdot  {\bf J}_O \over | {\bf J}_D||{\bf J}_O|} .
\end{displaymath}

\noindent Here, ${\bf J}_O$ is the orbital angular momentum. 
The disc angular momentum is ${\bf J}_D=\sum_j {\bf J}_j,$ where the
sum is over all disc particle angular momenta ${\bf J}_j.$
 
\noindent A precession angle $\beta_p,$ measured in the orbital plane 
can be defined through

\begin{displaymath}
\cos \beta_p = - { ({\bf J}_D \mbox{\boldmath $\times$} 
{\bf J}_O)\cdot {\bf u} \over | {\bf
J}_D{\mbox{\boldmath $\times$} \bf J}_O| |{\bf u}|} 
\end{displaymath}

\noindent where ${\bf u}$ may be taken to be any fixed reference 
vector in the orbital plane. We take this to point along the $y_o$
axis such that initially $\beta_p = \pi /2$ in all cases. For retrograde
precession of ${\bf J}_D$ about ${\bf J}_O$ the angle $\beta_p$ should
initially decrease as is found in practice. 

\noindent For a disc with constant $\Sigma$ and radius $R,$ 
the period of  rigid body precession of a thin disc is $2\pi /\omega_{p}$ where,
from equation~(\ref{PRECESS}) we obtain

\begin{equation}
\omega_p = - \left( {3GM_s \over 4D^3} \right) \cos \delta
{\int^R_0 \Sigma r^3 dr \over \int^R_0 \Sigma r^3 \Omega dr} 
= -{15 M_s R^3 \over 32 M_p D^3} \Omega(R)\cos \delta .
\label{Prc}
\end{equation}

\noindent  The condition for
 sound to propagate throughout the disc during a precession time
is approximately that

\begin{equation} 
\frac{H}{R} > \frac{|\omega_p|}{\Omega(R)} . 
\label{cond} 
\end{equation}

\noindent For models 1-3,
equation~(\ref{Prc}) gives $\omega_p/\Omega(R) = 0.012$ corresponding
to a precession period of 512  time units. Our results were
consistent with the condition~(\ref{cond}) to within a factor of two
in that models 1 and 2 with ${\cal M} < 25,$
showed modest warps and approximate rigid body precession while Model~3
with ${\cal M} = 30$ showed severe warping and a more complex
precessional behaviour.

\begin{figure}
%\centerline{\psfig{file=MACH20.D=3.t=0.ps,width=130mm}}
\caption{Projection plots in orbital plane coordinates for model~1
at time $t \simeq 0$. The projections are in the $(x,y_o)$ plane (top
left), $(x,z_o)$ plane (top right), $(y_o,z_o)$ plane (bottom left)
and $(x,z_o)$ plane (bottom right).}
\label{fig1}
\end{figure}
\newpage
\begin{figure}
%\centerline{\psfig{file=MACH20.D=3.angle.ps,width=130mm}}
\caption[]{The precession angle $\beta_p$ (dashed line) and inclination angle
$\iota$ (solid line) for model~1.}
\label{fig2}
\end{figure}

\begin{figure}
%\centerline{\psfig{file=MACH20.D=3.t=297.8.ps,width=130mm}}
\caption{Projection plots in orbital plane coordinates for model~1
at time $t=297.8$.}
\label{fig5}
\end{figure}
\begin{figure}
%\centerline{\psfig{file=fig7.ps,width=130mm}}
\caption{Projection plots in orbital plane coordinates for model~3
at time $t=397.9$.}
\label{fig7}
\end{figure}
\begin{figure}
%\centerline{\psfig{file=MACH30.D=3.angle.ps,width=130mm}}
\caption{The precession angle $\beta_p$ (solid and long-dashed lines)
and inclination angle $\iota$ (short-dashed and dotted lines) for
model~3. The solid and short-dashed lines correspond to the outer disc
section with $r>0.5$, and the long-dashed and dotted lines correspond
to the inner disc section with $r<0.3$.}
\label{fig8}
\end{figure}
\begin{figure}
%\centerline{\psfig{file=MACH25.D=3.angle.ps,width=130mm}}
\caption{The precession angle $\beta_p$ (solid and long-dashed lines)
and inclination angle $\iota$ (short-dashed and dotted lines) for
model~2. The solid and short-dashed lines correspond to the outer disc
section with $r>0.5$, and the long-dashed and dotted lines correspond
to the inner disc section with $r<0.3$.}
\label{fig8bis}
\end{figure}

\noindent We now
present particle projection plots for each of the Cartesian planes
using obital plane coordinates. In all such figures a fourth
`sectional plot' is also included in which only particles such that
$-0.05 < y_{o}< 0.05$ are plotted.

\subsection{Model 1 }
A projection plot for model~1 is shown in
Fig.~(\ref{fig1})
near $t=0$ when the disc is
unperturbed. Note that the disc appears as edge on and inclined at $45$
degrees in the $(y_o ,z_o)$ plane.
The time dependence of the angles $\iota$ and $\beta_p$ is plotted in
Fig.~(\ref{fig2}). It may be seen that there is little change in $\iota$ during
the whole run. On the other hand $\beta_p$ decreases approximately linearly,
corresponding to uniform precession (note that $\beta_p$ is plotted as positive
rather than negative for the latter section of this plot). The inferred
precession period is around 600 units, in reasonable agreement with the value of
512 units obtained from equation~(\ref{Prc}). 
\noindent Fig.~(\ref{fig5}) shows a projection plot
at $t=297.8$, near the end
of the run when the disc has precessed through about 180 degrees. This
amount of precession is demonstrated by the fact that the disc appears
almost edge on in the $(y_o ,z_o)$ plane just as it did at time $t=0.$ However,
its plane is inclined at about $90$ degrees to the original disc plane. At this stage
our results indicate that the disc has attained a quasi-steady configuration
as viewed in a frame that precesses uniformly about the orbital rotation axis.
The disc develops a warped structure that initially
grows in magnitude but then levels off.  The sectional plot
in Fig.~(\ref{fig5}) indicates that
the disc has developed a modest warp in this case.

\subsection{Models~2, and 3}
The behaviour of model~3  with ${\cal
M}=30$ 
is considerably more complex than that of model 1. This disc develops a strong 
warp such that the
inner and outer parts of the disc try to separate. A projection plot
is shown at $t=397.9$ in Fig.~(\ref{fig7}). The inner part of the disc
seems to occupy a different plane from the outer part. The outer part
was found to precess like a rigid body at the expected rate, and it
tended to drag the inner section behind it. This is indicated in
Fig.~(\ref{fig8}) where we plot the angle $\beta_p$ calculated using
the outer disc section with $r> 0.5$ only and also the same angle
calculated using  only the inner section with $r< 0.3.$ The angle associated
with the inner segment progresses at a variable rate indicating coupling to the
outer section. The angle $\iota$ associated with the two sections is also
plotted. This becomes significantly smaller for the inner segment consistent with
the existence of a large amplitude warp. We note that the relatively
larger inclination associated with the outer segment enables a closer
matching of the precession frequencies associated with the two
segments and aids coupling, due to the presence of large 
pressure gradients induced by the strong warping. 
Each segment of this model remained thin
throughout the run. 

\noindent  Fig.~(\ref{fig8bis}) shows the
evolution of $\beta_p$ and $\iota$ for model~2, which has ${\cal M}=25$ and is
 intermediate between model~1 and model~3. There is an indication of
differential precession initially.
But the inner part of the disc  couples to the outer part in such a way that the
precession becomes uniform after about 150 time units. It appears tha the inner
part is able to adjust its precession frequency to the outer part by  changing
slightly its relative inclination.  In this way the dependence of the precession
frequency on inclination is exploited to remove the differential precession.

\section{Discussion}
 
We have described   nonlinear simulations of an
accretion disc in a close  binary system when the disc
midplane is not necessarily coplanar with the plane of the binary
orbit. For our constant viscosity SPH models we found  the  tidal
truncation phenomenon to be only marginally affected by non
coplanarity. We found that modestly warped and thin discs
undergoing near rigid body precession may survive in close binary
systems. However, extremely thin discs may be severely disrupted by
differential precession  depending  on the
magnitude of the characteristic  Mach number, ${\cal M}.$
The crossover
between obtaining a warped, but coherent disc structure, and disc disruption
 occurs  for a value of the Mach number ${\cal M} \sim 30.$ We also
found that the inclination evolved on a long timescale, likely to be the viscous
timescale, as indicated by the linear calculations of Papaloizou and
Terquem (1995).

\noindent A class of
models  formulated to explain the
generation of jets in young stellar objects assumes that a wind flows outwards
from the disc surface. This is then accelerated and collimated by the action of
a magnetic field (see  K\"onigl and  Ruden, 1993 and references therein).
It is reasonable to assume that  a precessing disc may
lead to the excitation of a precessing jet. The precession period
obtained from our calculations with a mass ratio of unity is about 500
 units. When scaled to a disc of radius 50 AU,
surrounding a star of $1 {\rm M}_{\odot}$, the unit of time is
$\Omega(R)^{-1} \simeq 56$~$yr$, leading to a precession period of
 $3.10^4 yr$.

\noindent Bally and  Devine (1994) suggest that the jet which seems to be
excited by the young stellar object HH34* in the L1641 molecular cloud
in Orion precesses with a period of approximately $10^4$~$yr$. This
period is consistent with the source being a binary with parameters
 similar to those we have used in our simulations with a separation on the
order of a few hundred astronomical units.
 
\noindent Some of the results presented here demonstrate how a warped
disc can present a large  surface area for intercepting the primary
star's radiation. The effect that the consequent reprocessing of the stellar
radiation field can have on the emitted spectral energy distribution
has been investigated by Terquem and  Bertout (1993, 1996). They find
that it  may account for the high spectral index of some T~Tauri stars or
even for the spectral energy distribution of some class~0/I
sources. The Model~3 simulation indicates that the required strongly
warped disc
could  be physically realisable.

\noindent Finally, there is evidence from the light curves of X--ray binaries
such as Hercules X-1 and SS433, that their associated accretion discs may be
 precessing in the tidal field
of the binary companion ( Schandl and  Meyer, 1994).  Larwood et al (1996)
have demonstrated that the disc precession periods seen in  simulations
are in reasonable agreement with those that are inferred observationally
( Petterson, 1975, Gerend and  Boynton, 1976, Margon, 1984).

\vspace{.2cm}\noindent{\it Acknowledgement:}
This work was supported by PPARC grant GR/H/09454, JDL is
supported by a PPARC studentship.


\begin{thebibliography}{}

\bibitem{}{1994A}{Artymowicz}
Artymowicz P., Lubow S.H., 1994, ApJ, 421, 651 

\bibitem{}{1994B}{Bally}
Bally J., Devine D., 1994, ApJ, 428, L65 

\bibitem{}{1992Bi}{Bibo}
Bibo E.A., The P.S., Dawanas D.N., 1992, A\&A, 260, 293 

\bibitem{}{1996C}{Corporon}
Corporon P., Lagrange A.M., Beust H., 1996, A\&A, 310, 228 

\bibitem{}{1987E}{Edwards}
Edwards S., Cabrit S., Strom S.E., Heyer I., Strom K.M., Anderson E.,
1987, ApJ, 321, 473 

\bibitem{}{1976G}{Gerend}
Gerend D., Boynton P., 1976, ApJ, 209, 562 

\bibitem{}{1977G1}{Gingold1}
Gingold R.A., Monaghan J.J., 1977, MNRAS, 181,375 

\bibitem{}{1981G}{Goldreich}
Goldreich P., Tremaine S., 1981, ApJ, 243, 1062 

\bibitem{}{1969Hu}{Hu}
Hunter, C. \& Toomre, A. 1969, {\it Ap. J.}, {\bf 155}, 747

\bibitem{}{1993Kon}{Konigl}
K\"onigl A., Ruden S.P., 1993, in Levy E.H., Lunine J., eds,
Protostars and Planets III (Univ. Arizona Press, Tucson), p. 641 

\bibitem{}{1996La}{LNPT}
Larwood J.D., Nelson R.P., Papaloizou J.C.B., Terquem C., 1996, MNRAS, 282, 597 
 
\bibitem{}{1997La}{Larwood}
Larwood J.D., Papaloizou J.C.B., 1997, MNRAS, In press 

\bibitem{}{1993Le}{Leinert}
Leinert C., Zinnecker H., Weitzel N., Christou J., Ridgway S.T.,
Jameson R.F., Haas M., Lenzen R., 1993, A\&A, 278, 129 

\bibitem{}{1993Li}{Lin}
Lin D.N.C., Papaloizou J.C.B., 1993, in Levy E.H., Lunine J., eds,
Protostars and Planets III (Univ. Arizona Press, Tucson) p.749 

\bibitem{}{1977Lu}{Lucy}
Lucy L.B., 1977, AJ, 83, 1013 

\bibitem{}{1967Lb}{Lynden1}
Lynden-Bell D., Ostriker J.P., 1967, MNRAS, 136, 293 

\bibitem{}{1967Lb}{Lynden2}
Lynden-Bell D., Pringle J.E., 1974, MNRAS, 168, 603 

\bibitem{}{1984Ma}{Margon}
Margon B., 1984, ARA\&A, 22, 507 

\bibitem{}{1994Mat}{Mathieu}
Mathieu R.D., 1994, ARA\&A, 32, 465 

\bibitem{}{1983Mo}{Monaghan}
Monaghan J.J., Gingold R.A., 1983, J. Comp. Phys., 52, 374 

\bibitem{}{1985Mo}{Monaghan}
Monaghan J.J., Lattanzio J.C., 1985, A\&A, 149, 135 

\bibitem{}{1993Ne}{Nelson}
Nelson R.P., Papaloizou J.C.B., 1993, MNRAS, 265, 905 

\bibitem{}{1994Ne}{Nelson}
Nelson R.P., Papaloizou J.C.B., 1994, MNRAS, 270, 1 

\bibitem{}{1994Neb}{Nelson}
Nelson R.P., 1994, Ph.D Thesis, University of London 
 
\bibitem{}{1995Pap2}{Papaloizou2}
Papaloizou J.C.B., Lin D.N.C., 1995, ApJ, 438, 841 

\bibitem{}{1993Pap3}{Papaloizou3}
Papaloizou J.C.B, Pringle J.E., 1983, MNRAS, 202, 1181 

\bibitem{}{1995Pap4}{Papaloizou4}
Papaloizou J.C.B., Terquem C., 1995, MNRAS, 274, 987 

\bibitem{}{1975Pet}{Petterson}
Petterson J.A., 1975, ApJ, 201, L61 

\bibitem{}{1994Sa}{Savonije}
Savonije G.J., Papaloizou J.C.B., Lin D.N.C., 1994, MNRAS, 268, 13 
 
\bibitem{}{1973sh}{Shakura}
Shakura N.I., Sunyaev R.A., 1973, A\&A, 24, 337 

\bibitem{}{1994SC}{Schandl}
Schandl S., Meyer F., 1994, A\&A, 289, 149

\bibitem{}{1984Sh}{Shu}
Shu, F. S. 1984. In Planetary Rings, Greenberg, R. \& Brahic, A., eds.,
University of Arizona Press.

\bibitem{}{1987Sp}{Spruit}
Spruit H.C., 1987, A\&A, 184, 173 

\bibitem{}{1993T1}{Terquem1}
Terquem C., Bertout C., 1993, A\&A, 274, 291 

\bibitem{}{1994T2}{Terquem2}
Terquem C., Bertout C., 1996, MNRAS, 279, 415 
 
\end{thebibliography}
 \end{document}